\documentclass[final,5p,times,twocolumn]{elsarticle}

\usepackage{graphicx}

\usepackage{amssymb}

\journal{Nuclear Instruments and Methods A}

\begin{document}

\begin{frontmatter}



\title{Development of Fast High-Resolution Muon Drift-Tube Detectors for High
Counting Rates}


\author[A]{B.~Bittner}
\author[A]{J.~Dubbert}
\author[A]{S.~Horvat}
\author[A]{O.~Kortner}
\author[A]{H.~Kroha}
\author[A]{F.~Legger}
\author[A]{R.~Richter}
\author[B]{S.~Adomeit}
\author[B]{O.~Biebel}
\author[B]{A.~Engl}
\author[B]{R.~Hertenberger}
\author[B]{F.~Rauscher}
\author[B]{A.~Zibell}
\address[A]{Max-Planck-Institut f\"ur Physik, F\"ohringer Ring 6, D-80805 Munich, Germany}
\address[B]{Ludwig-Maximilians-Universit\"at M\"unchen, Am Coulombwall 1, D-85748 Garching, Germany}

\begin{abstract}

Pressurized drift-tube chambers are efficient detectors for high-precision
tracking over large areas. The Monitored Drift-Tube (MDT)
chambers of the muon spectrometer of the ATLAS detector at the Large Hadron
Collider (LHC) reach a spatial resolution of 35~$\mu$m and almost $100\%$ tracking efficiency
with 6 layers of 30~mm diameter drift tubes operated with Ar:CO$_2$ (93:7) 
gas mixture at 3 bar and a gas gain of 20000. The ATLAS MDT chambers are 
designed to cope with background counting rates due to neutrons and $\gamma$ rays of up to
about 300~kHz per tube which will be exceeded for LHC luminosities larger than the design value
of $10^{34}$~cm$^{-1}$~s$^{-1}$. 
Decreasing the drift-tube diameter to 15~mm while keeping the other parameters,
including the gas gain, unchanged reduces the maximum drift time from about 700~ns to 200~ns
and the drift-tube occupancy by a factor of 7. New drift-tube chambers for the endcap regions of the ATLAS
muon spectrometer have been designed. A prototype chamber consisting of 12 times 8 layers of
15~mm diameter drift tubes of 1~m length has been constructed with a sense wire positioning accuracy of 
20~$\mu$m. The 15~mm diameter drift-tubes have been tested with cosmic rays
in the Gamma Irradiation Facility at CERN at $\gamma$ counting rates of up to 1.85~MHz.

\end{abstract}

\begin{keyword}

Muon chambers \sep drift tubes \sep high rates \sep LHC


\end{keyword}

\end{frontmatter}


\section{Introduction}

The muon detectors of the experiments at the Large Hadron Collider (LHC) 
will encounter unprecedentedly high background counting rates due to neutrons and $\gamma$ rays
in the energy range up to about 10~MeV which originate mainly from secondary interactions
of the hadronic collision products with accelerator elements, shielding material and the
detector components. The LHC schedule foresees a continuous increase of the luminosity 
eventually exceeding the original design value of $10^{34}$~cm$^{-2}$~s$^{-1}$.
Assuming that the background rates approximately scale with the luminosity,
the rate capability of the muon detectors will be exceeded.
The Monitored Drift-Tube (MDT) chambers in the muon spectrometer 
of the ATLAS detector at the LHC~\cite{MuonTDR,ATLASpaper}, for example, are designed to cope with
counting rates of up to about 300~kHz in the endcap regions of the spectrometer
corresponding to an occupancy of $21\%$.

The MDT chambers consist of two triple or quadruple layers (multilayers) of
aluminum drift tubes of 30~mm outer diameter and 0.4~mm wall thickness filled with Ar:CO$_2$ (93:7) gas mixture at
a pressure of 3~bar. A voltage of 3080~V is applied between the tube wall and the $50~\mu$m diameter anode wire
corresponding to a gas gain of $2\cdot 10^4$ at low counting rates and resulting in a maximum drift time of about 700~ns. 
At low counting rates, the average spatial resolution of individual drift
tubes is $80~\mu$m applying time-slewing corrections based on pulse height measurement~\cite{Horvat}. With a sense-wire
positioning accuracy of better than $20~\mu$m, this corresponds to a chamber spatial resolution of 35~$\mu$m.

We investigate the possibility of using pressurized drift-tube detectors
with smaller tube diameter and therefore shorter maximum drift-time in the highest background regions
of the muon detectors of the LHC experiments. Building on the experience with the ATLAS MDT chambers,
new muon drift-tube detectors with 15~mm diameter tubes have been developed 
which can cope with 10 times higher background fluxes.

\begin{figure}[!ht]
\centering
\includegraphics[width=0.45\textwidth,keepaspectratio]{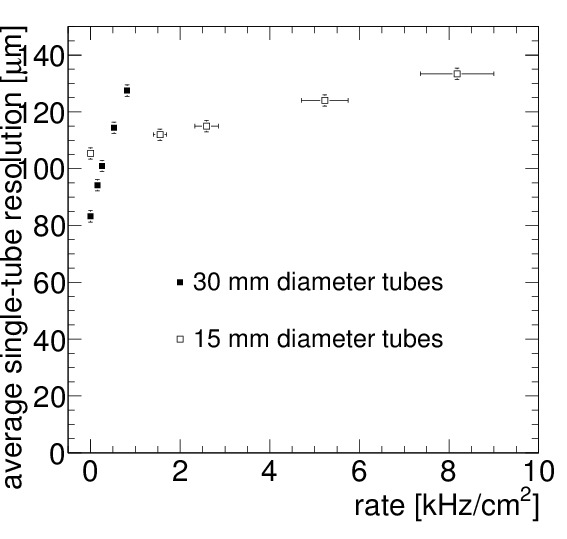}
\caption{Average spatial drift-tube resolution as a function of the background counting rate in 15 mm and 30 mm diameter tubes.}
\label{fig:resol}
\end{figure}

\begin{figure*}[!ht]
\centering
\includegraphics[width=\textwidth,keepaspectratio]{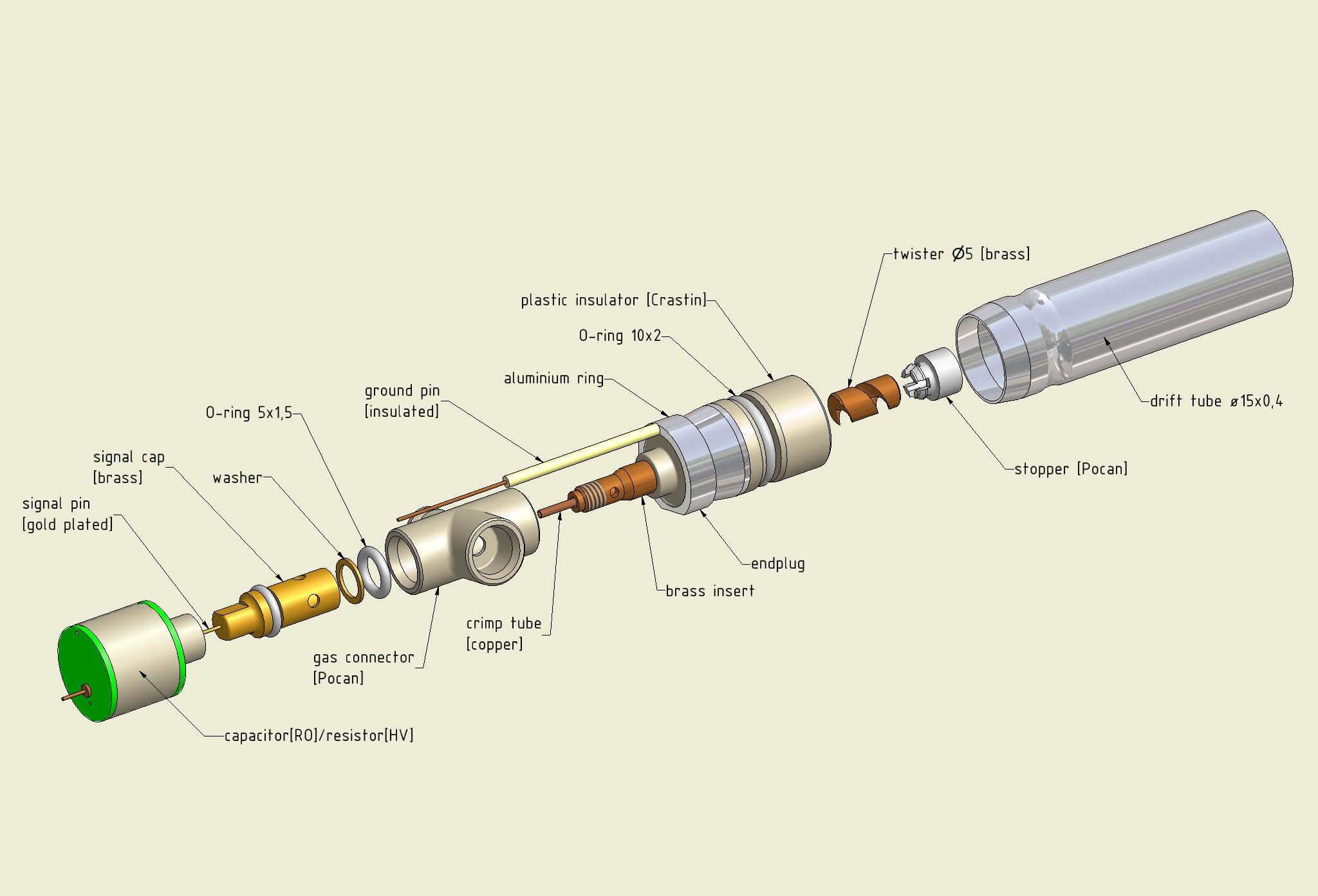}
\caption{Exploded view of the endplug for 15~mm diameter drift tubes with interfaces to the gas distribution 
and the electronics boards. The spiral-shaped wire locator (twister) with $50~\mu$m innner diameter fits into the 
central bore of the brass insert (the arrow points to the external reference surface for precise wire positioning
in the chamber).}
\label{fig:endplug}
\end{figure*}

\section{Drift-tube performance at high rates}

At high counting rates, the drift tubes of the ATLAS MDT chambers are known to suffer
from a degradation of the spatial resolution due to space-charge effects~\cite{Aleksa,Deile}
and of the muon detection efficiency due to the increased drift-tube occupancy~\cite{Horvat}.
Both effects can be supressed by reducing the tube diameter while leaving the other operating
parameters of the drift tubes, in particular gas mixture, pressure and gas gain, unchanged.

Decreasing the drift-tube diameter from 30~mm to 15~mm and the operating voltage from 3080 to 2730~V
leads to a reduction of the maximum drift time by a factor of 3.5 from about 700~ns to 200~ns~\cite{Bittner}.
In addition, the background counting rate, dominated by the conversion of the neutron and gamma radiation 
in the tube walls, decreases proportional to the tube diameter, i.e.\ by a factor of two per unit tube length.
Both effects together lead to a reduction of the occupancy by about a factor of 7.
At the same time, at least twice the number of drift-tube layers can be accommodated in the same
detector volume allowing for additional improvement of the muon detection efficiency and spatial resolution.

At high counting rates, the space-charge generated by the ion clouds drifting towards the tube wall
lowers the effective potential near the anode wire leading to a reduction of the gas gain.
The resulting loss in signal height and, therefore, spatial resolution 
grows with the inner tube radius $R_2$ proportional to $R_2^3\cdot \ln (R_2/R_1)$~\cite{Riegler} where $R_1=25~\mu$m
is the wire radius. Therefore, the signal height reduction due to space charge is 10 times smaller in 15 mm 
compared to 30 mm diameter tubes.
Fluctuations of the space charge and, consequently, of the electric field in the tube lead to variations of the drift velocity
in non-linear drift gases like Ar:CO$_2$ (93:7) causing a deterioration of the spatial resolution which
increases rapidly with the drift distance above a value of about 7.5~mm~\cite{Aleksa,Deile}. In addition, 
the space-to-drift time relationship for the Ar:CO$_2$ (93:7) drift gas is more linear at drift distances
below 7.5~mm reducing the sensitivity of the position measurement to environmental parameters such as gas composition
and density, magnetic field and, in particular, irradiation rate.

Since the drift-tube spatial resolution at low background rates improves with the drift distance, the average single-tube
resolution deteriorates from $80~\mu$m for 30~mm diameter tubes~\cite{Horvat, Deile} to about $110~\mu$m
for 15~mm diameter tubes. For 30~mm diameter drift-tubes, the
resolution has been measured to deteriorate appproximately linearly with the counting rate to about $120~\mu$m at
500~Hz/cm$^2$~\cite{Horvat, Deile}. For 15~mm diameter tubes, the rate dependence of the resolution, 
dominated by the gain drop effect, is expected to be about 10 times smaller (see Fig.~\ref{fig:resol}).

\begin{figure}[!ht]
\centering
\includegraphics[width=0.3\textwidth,keepaspectratio]{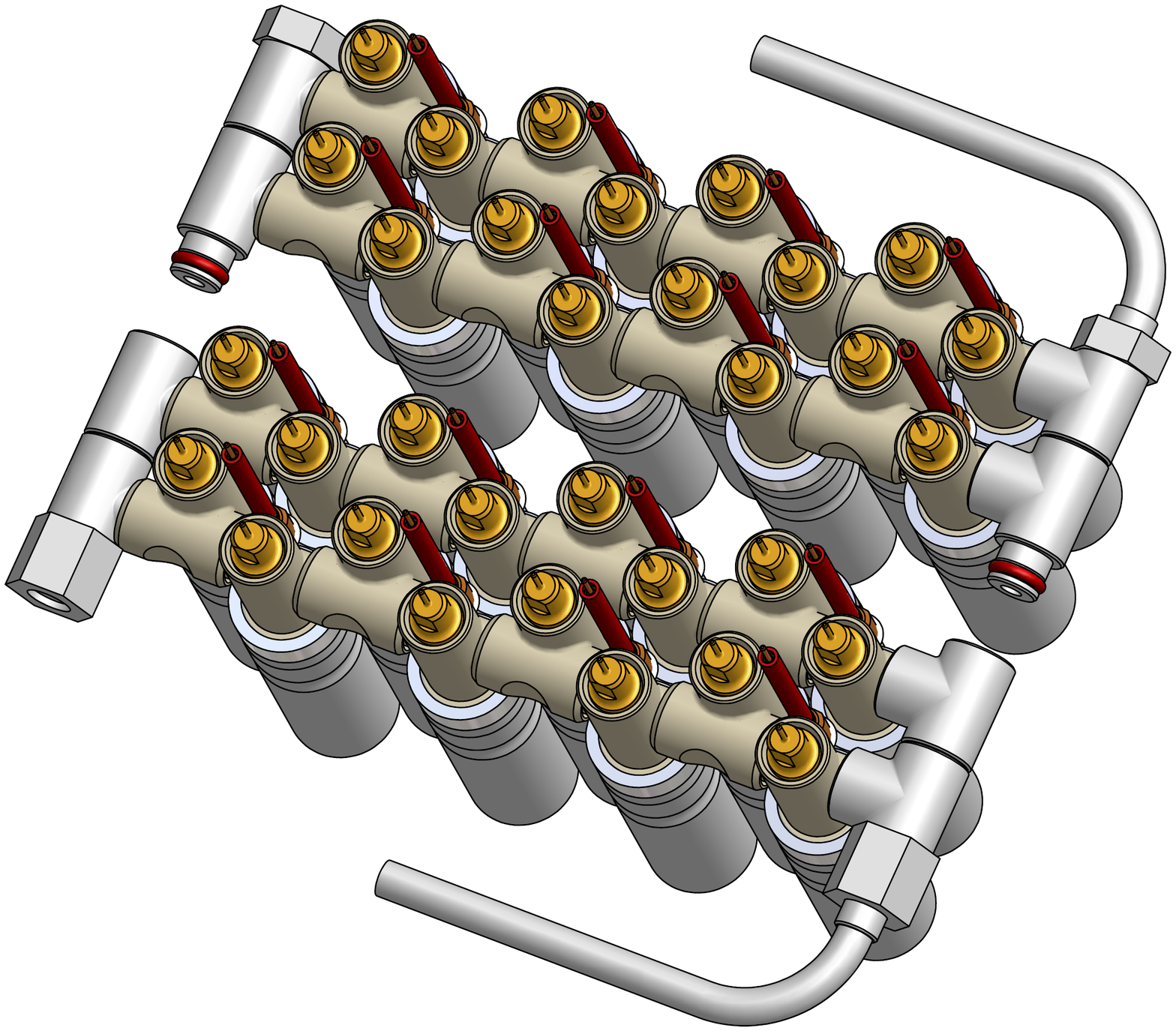}
\caption{Modular gas distribution system consisting of injection moulded plastic connections
between the tubes and to the gas supply pipes.}
\label{fig:gassystem}
\end{figure}

\begin{figure}[!ht]
\centering
\includegraphics[width=0.35\textwidth,keepaspectratio]{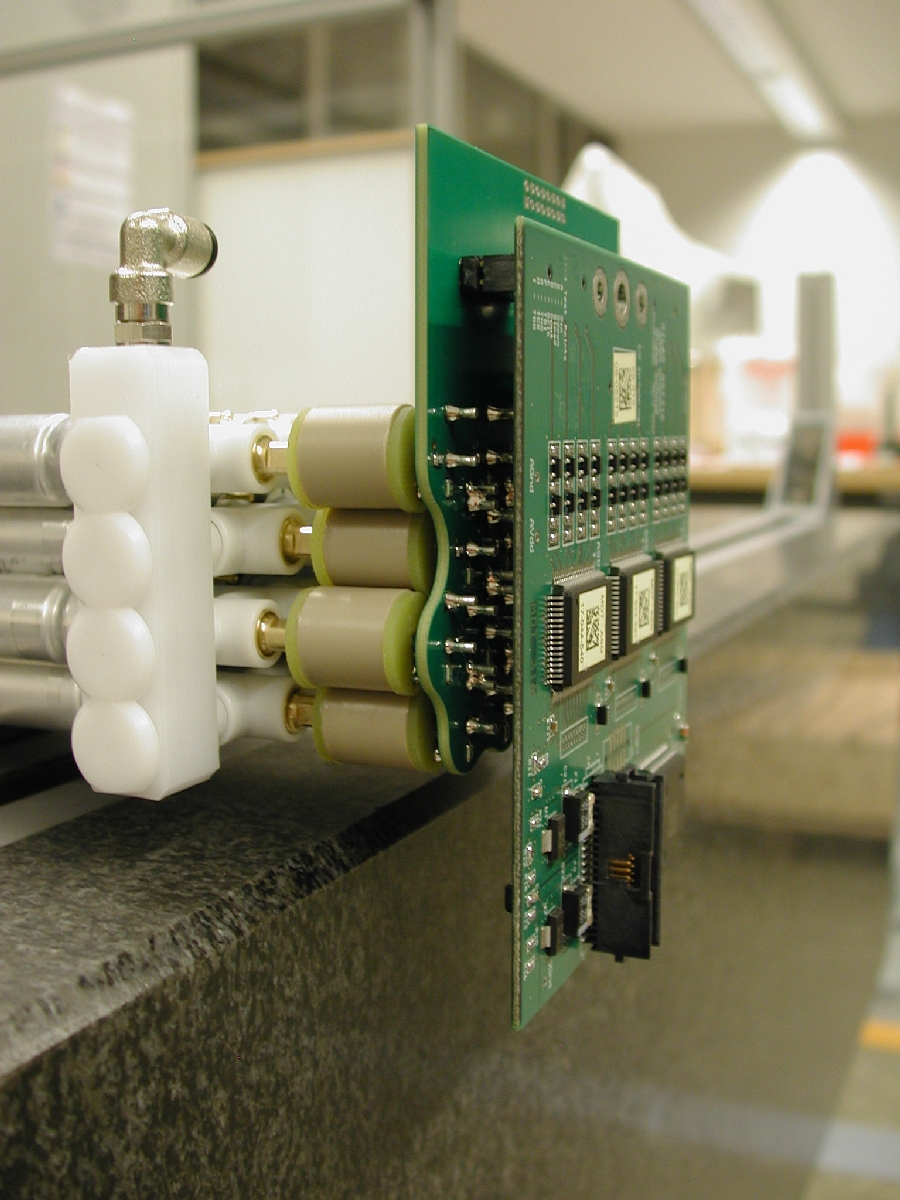}
\caption{Photograph of the interface between the drift tubes and the readout electronics.
The white plastic gas connectors and the insulating cans for the high-voltage
decoupling capacitors between the signal caps on the endplugs and the readout boards are visible.}
\label{fig:interface}
\end{figure}

\begin{figure}[!ht]
\centering
\includegraphics[width=0.45\textwidth,keepaspectratio]{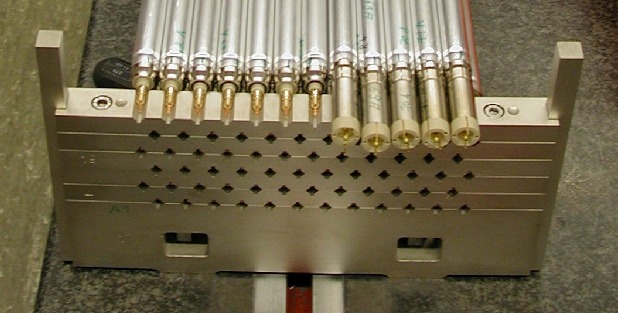}
\caption{Assembly of a prototype chamber with 15~mm diameter drift tubes. 
The external endplug reference surfaces at both tube ends are inserted in precise mechanical jigs 
(before the installation of gas connections and signal caps) in order to position the sense wires
with better than $20~\mu$m accuracy.}
\label{fig:module}
\end{figure}

\section{Chamber design and fabrication}

Central to the chamber design is the development of endplugs for the 15~mm diameter tubes (see Fig.~\ref{fig:endplug})
which builds on the endplug design for the ATLAS MDT chambers~\cite{Bauer}.
The injection moulded plastic endplugs electrically insulate the sense wire from the tube wall and
center the wire in the tube with few micron accuracy with respect to the external reference surfaces on the central
brass inserts of the endplugs. The wire is tensioned and fixed in copper crimp tubelets at the ends of the brass inserts.
The endplugs also provide the interfaces to the gas distribution system (see Fig.~\ref{fig:gassystem} and to the readout
and high-voltage distribution boards (see Fig.~\ref{fig:interface}). Ground pins inserted between adjacent tubes electrically
interconnect the tube walls and connect them to ground. The decoupling capacitors on the readout end of the tubes and the
protection resistors on the high-voltage connection side are housed in insulating plastic cans. 

The tubes are assembled to a chamber using precise mechanical jigs (see Fig.~\ref{fig:module} positioning the sense wires
relative to each other with better than $20~\mu$m accuracy using the external reference surfaces of the endplugs.
All tubes of a multilayer can be assembled and glued together in a single step requiring only one working day.
A prototype chamber consisting of 12 layers with eight one meter long drift tubes has been constructed (see Fig.~\ref{fig:module}).
Standard aluminum tubes with 0.4~mm wall thickness and tolerances of $\pm 0.1$~mm on diameter, roundness
and concentricity and of $\pm 0.5$~mm on straightness have been used successfully. Measurements of the prototype chamber in a cosmic
ray test stand with two ATLAS MDT chambers as tracking references~\cite{Biebel} confirmed the desired wire positioning accuracy 
of $20~\mu$m.

\section{Test results}

\begin{figure}[htb]
\centering
\includegraphics[width=0.45\textwidth,keepaspectratio]{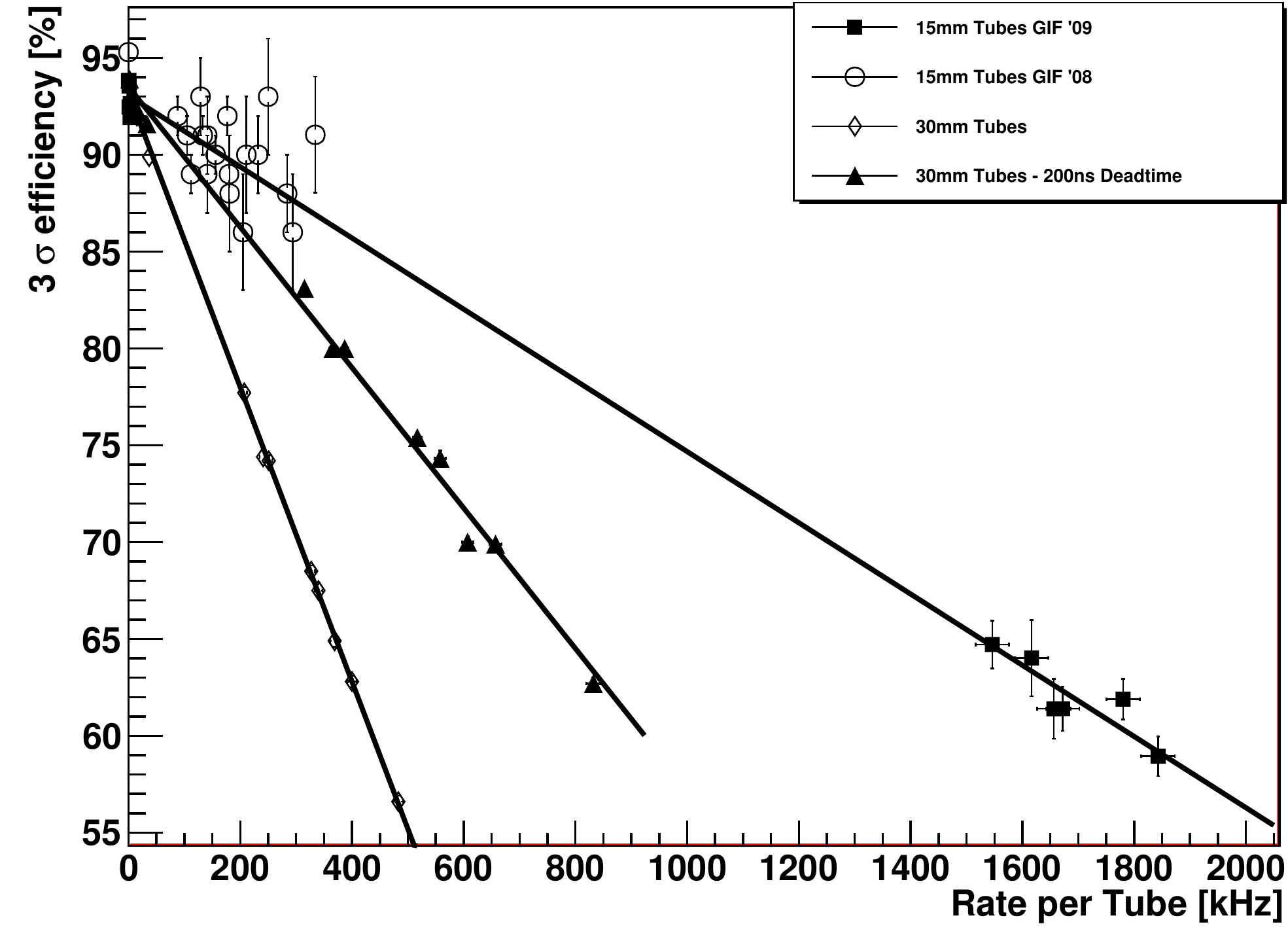}
\caption{Muon detection efficiency (see text) of 15 and 30~mm diameter drift tubes as a function of the
$\gamma$ ray background counting rate.}
\label{fig:efficiency}
\end{figure}

Drift tubes of 15~mm diameter and one meter length equipped with the ATLAS MDT readout electronics~\cite{ATLASpaper}
have been tested with cosmic ray muons in the Gamma Irradiation Facilty (GIF) at CERN at $\gamma$ counting rates of 
up to 1.85~kHz.
The test setup with two MDT reference chambers above and below the tested detectors is described in~\cite{Dubbert}.
The adjustable deadtime of the readout electronics was set to its minimum value of 200~ns.
With increasing background counting rate, the muon hits are increasingly masked by background hits. The measured probability
to detect a muon hit at the drift radius expected within three times the spatial drift-tube resolution of the
extrapolation from the reference chambers is shown in Fig.~\ref{fig:efficiency} as a function of the counting rate
in comparison with previous measurements for 30~mm diameter drift tubes with the same readout electronics
with 200 and 790~ns deadtime setting~\cite{Horvat}. The drift-tube efficiency is considerably increased as expected, approximately by the ratio of the sums of the maximum drift time and the electronics dead time.

\end{document}